# Fluid-Based Analysis of Pedestrian Crowd at Bottlenecks

**Peng N. Wang** **Peter B. Luh**

*Abstract—*

**In emergency egress crowd behavior critically affects egress efficiency and public safety. By integrating psychological principles to Newtonian motion of crowd, a fluid-based equation is derived in this paper to explore how energy in different forms is balanced when pedestrian crowd pass through a bottleneck. Such fluid-based analysis helps to bridge a gap among psychological findings, pedestrian models and simulation results, and it further provides a new perspective to understand how the faster-is-slower effect is caused and how disastrous events (e.g., jamming and stampede) occur at a bottleneck passage.**

## I. Introduction

"People kept pushing down into a downhill club alley, resulting in other people screaming and falling down like dominos," an unidentified witness wrote about the deadliest stampede in South Korea's history. Such a tragedy happened in the famous nightlife district of Itaewon after tens of thousands of people visited the area for Halloween, and a total of 151 people were killed and 82 injured in the horrible moments of pushing, shoving and crushing. Similar lessons have been learned many times historically (Still, 2016), and the fundamental cause of such disasters seems unpredictable and unclear to the scientific circle, and how to effectively mitigate or prevent such disasters is important.

To well understand critical behavior in crowd disaster, simulation study has been widely applied, in which an individual's behavioral and psychological status is microscopically modeled by a set of mathematical equations or logic rules, and their interaction and collective behavior are calculated by numerical methods. This approach was rapidly boosted with advanced computing technology at end of last century. The mathematical equations of each individual is mainly inspired from physics, such as cellular automaton model, lattice gas model and force-based model (Zheng et. al., 2009). A well-known model is called social force model, which describes psychological features such as impatience and neverousness of individuals, and the simulation results demonstrated several paradoxical phenomenon between individual psychological variables and physical measurement of crowd, such as "faster-is-slower" and "freezing-by-heating" effect. In sum the microscopic models enable us to take into account a variety of pedestrian characteristics, and we can observe the complex nonlinear dynamics in the simulation result. However, sometimes we lack analytical tools to deeply understand what we have observed in simulation. Thus, we also need the second method at the macroscopic level.

The second approach refers to an analytical model at the macroscopic level, aggregating many particle-like individuals directly into fluid-like model of crowd, and we call it fluid-based analysis of pedestrian crowd. In other words the crowd fluid model (macro-level) is derived from the individual motion equation (micro-level) based on physics laws and mathematical principles. The resulting model are a set of partial differential equation (PDE), which could be dated back to

Peng Wang previously studied in the Department of Electrical and Computer Engineering, University of Connecticut, Storrs, USA, Email: wp2204@gmail.com

the well-known Eular Equation or Navior-Stocks Equations in 18th and 19th centuries. Since 1950s the equations are modified in study of road traffic problem (Lighthill and Whitham, 1955), and further extended to pedestrian traffic problem (Al-nasur and Kachroo, 2006). The major merits of the approach is that the model itself has long been established in physics, and there has been a number of analytical and numerical tools, providing valuable insights to understand observation in micro-simulation as well as real-world phenomena. However, analog of fluid dynamics is only valid for high-density crowd. If the crowd density is low, the continuity hypothesis fails and the method cannot be applied.

Although the methods could be either microscopic or macroscopic, they are different paths to the same problem, exploring the same phenomenon of crowd from different perspectives. Thus, advances in micro-models also facilitate model at macro-level, and verse visa. Based on this logic, this article is partly inspired from the microscopic model (e.g, social force model) such that we introduce psychological variables from the micro-level analysis to fluid-based analysis of high-density crowd at bottlenecks (e.g, narrow doorways). The resulting equations essentially describe the effect of perceived stress on crowd speed and density in terms of energy conservation: the psychological drive arises as a form of potential energy, which can be transformed to either a kinetic form (i.e., speeding up crowd movement), or to a static form (i.e., increasing crowd density). When crowd motion is limited by passage capacity, the energy will be transformed to a static form instead of kinetic form, resulting in an increase of crowd density. High crowd density at narrow passages could intensify physical interaction among people, resulting in the disastrous events of disorder and stampede.

The crowd fluid dynamics reflects a common psychological finding: moderate stress improves human performance (i.e., reducing the egress time); while excessive stress impairs it (i.e., increasing crowd density and causing disorder and blocking at bottlenecks). The model is not only psychologically meaningful – it provides insight to understanding the model and simulations of Helbing, Farkas and Vicsek, 2000 from a psychological perspective – but it is also practical: it abstracts crowd motion at a macroscopic level so that it forms a basis for analysis and optimization of crowd evacuation in the future (Wang et. al., 2008). As an integration of multiple disciplines, this model advances cross-disciplinary knowledge of psychology, egress modeling and engineering studies, and it holds promise of long-term impact on egress analysis by exploring transformative optimization methods for time-critical evacuation events.

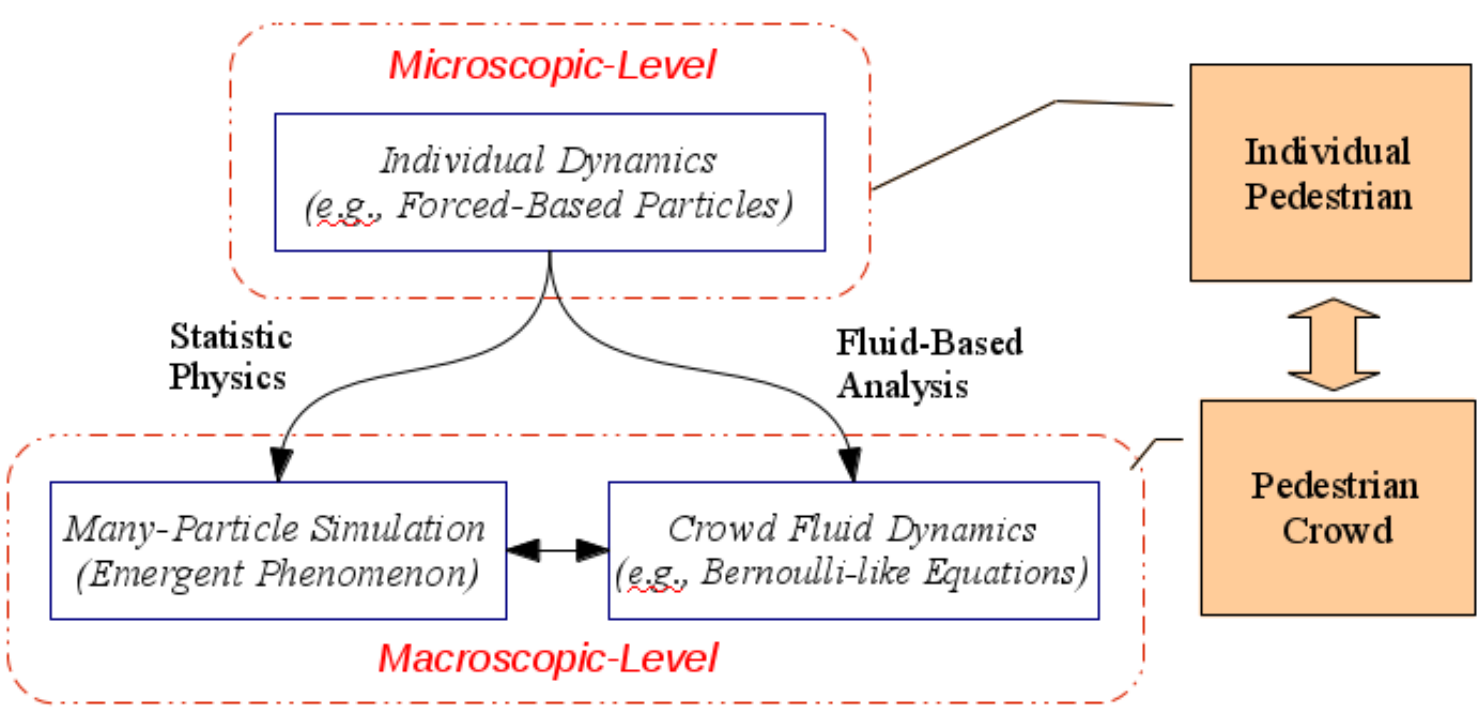

**Figure 1. Fluid-Based Analysis of Crowd Movement: The method is generally from fluid physics, which studies how individual agent interacts and emerge certain characteristics at the macroscopic level such as crowd speed and density.**

## II. Review of the social-force Model and Simulations

First, let us briefly review the force-based description of the microscopic pedestrian model. This model assumes that an individual's motion is motivated by a driving force $\boldsymbol{f}_i^{self}$ and the resistances come from surrounding individuals and facilities. Let $\boldsymbol{f}_{ij}$ denote the interaction from individual $j$ to individual $i$, and $\boldsymbol{f}_{iw}$ denote the force from walls or other facilities to individual $i$. Given the instantaneous velocity $\boldsymbol{v}_i(t)$ of individual $i$, the moving dynamics is given by the Newton Second Law:

$$m_i \frac{d v_i(t)}{\mathrm{d}t} = f_i^{drv} + \sum_{j(\neq i)} f_{ij} + \sum_{w} f_{iw} + \xi_i \qquad (1)$$

where $m_i$ is the mass of individual $i$ and $\boldsymbol{v}_i(t)$ is its moving velocity at time $t$. $\boldsymbol{\xi_i}$ is a small fluctuation force, resulting in random motion, and it indicates the "heating effect" on the crowd. Furthermore, the self-driving force is specified by

$$f_i^{self} = \frac{m_i}{\tau_i}\left(v_i^0(t) - v_i(t)\right), \tag{2}$$

The above driving force describes an individual tends to move with a certain desired velocity $\boldsymbol{v}_i^0(t)$ and expects to adapt the instantaneous velocity $\boldsymbol{v}_i(t)$ within a certain time period $\tau_i$. In particular, the desired velocity $\boldsymbol{v}_i^0(t)$ is the target velocity in one's mind, specifying the speed and direction that one expects to realize. The physical velocity $\boldsymbol{v}_i(t)$ is the physical speed and direction achieved in reality. The gap between $\boldsymbol{v}_i^0(t)$ and $\boldsymbol{v}_i(t)$ implies the difference between the subjective wish in people's mind and objective situation in the physical reality, and it is scaled by a time parameter $\tau_i$ to form the driving force. As a result, the driving force contributes to either acceleration or deceleration, and it functions in a feedback loop in a pedestrian dynamics, making the realistic velocity $\boldsymbol{v}_i(t)$ approaching towards the desired velocity $\boldsymbol{v}_i^0(t)$ in most situations. This linear form of driving force could be dated back to the Payne-Whitham traffic flow model (Payne, 1971; Whitham, 1974), where the desired velocity is defined as the flow velocity in equilibrium states.

The interaction force of pedestrians consists of two parts: the social force $\boldsymbol{f}_{ij}^{soc}$ and physical interaction $\boldsymbol{f}_{ij}^{phy}$ .i.e., $\boldsymbol{f}_{ij} = \boldsymbol{f}_{ij}^{soc} + \boldsymbol{f}_{ij}^{phy}$. The social force $\boldsymbol{f}_{ij}^{soc}$ was initially introduced in Helbing and Molnar, 1995 as an exponential function of distance between individuals $d_{ij}$. The force increases with reducing distance $d_{ij}$ , characterizing the social-psychological tendency of two pedestrians to keep proper interpersonal distance. In Wang, 2016 the social force was modified by introducing a new concept of desired interpersonal distance $d_{ij}^0$, and $\boldsymbol{f}_{ij}^{soc}$ is generally a function of $d_{ij}^0 - d_{ij}$ , namely $\boldsymbol{f}_{ij}^{soc} = \boldsymbol{F}_{ij}(d_{ij}^0 - d_{ij})$, where $\boldsymbol{F}_{ij}()$ represents a vectorial function pointing from pedestrian $j$ to pedestrian $i$. Following the exponential form of social force in Helbing and Molnar, 1995, we introduce the modified social force by $A_i exp((d_{ij}^0 - d_{ij})/B_i)\boldsymbol{n}_{ij}$, where $d_{ij}^0$ is the counter part of desired velocity $\boldsymbol{v}_i^0(t)$ in Equation (2).

The physical force $\boldsymbol{f}_{ij}^{phy}$ describes the physical interaction when pedestrians have body contact. The physical interaction $\boldsymbol{f}_{ij}^{phy}$ is composed by an elastic force that counteracts body compression and a sliding friction force that impedes relative tangential motion of two pedestrians. The interaction of a pedestrian with obstacles like walls is denoted by $\boldsymbol{f}_{iw}$ and is treated analogously. In Helbing, Farkas and Vicsek, 2000 the interaction forces among individuals are repulsive. The model may also include an attraction force in its original version (Helbing and Molnar, 1995, Korhonen and Hostikka, NIST, 2010). By simulating many such individuals in collective motion, blocking was observed as they pass a bottleneck doorway. This scenario is named by the "faster-is-slower" effect in Helbing, Farkas and Vicsek, 2000. In particular, it shows that increasing psychological state (i.e., desired velocity $\boldsymbol{v}_i^0$) can inversely decrease physical collective speed of crowd passing through the doorway.

In the above model we emphasize that the social-force and driving force are not physics-based concepts because they are psychologically generated from people's conscious mind. The driving force represents a motive of behavior and it relates to time-pressure of environment or impatience of individuals. The social force reflects a kind of space-related stress, describing how people use their personal space to interact with surrounding people. Both of the forces are realized by foot-and-ground friction in the physical world. The foot-and-ground friction is the physics concept, and we are able to consciously adjust or control the friction to decide where and how fast we move (Wang, 2016). A major assumption in the force-based model refers to vectorial additivity of force components, which implies different mental processes could function in parallel, and are formalized in a linear form. This assumption is based on the psychological facts that human mind has the capability of parallel perceiving and thinking, which means people are able to adjust their interpersonal distance (i.e., social force) while in the meantime still keep their destination in mind (i.e., self-driving force).

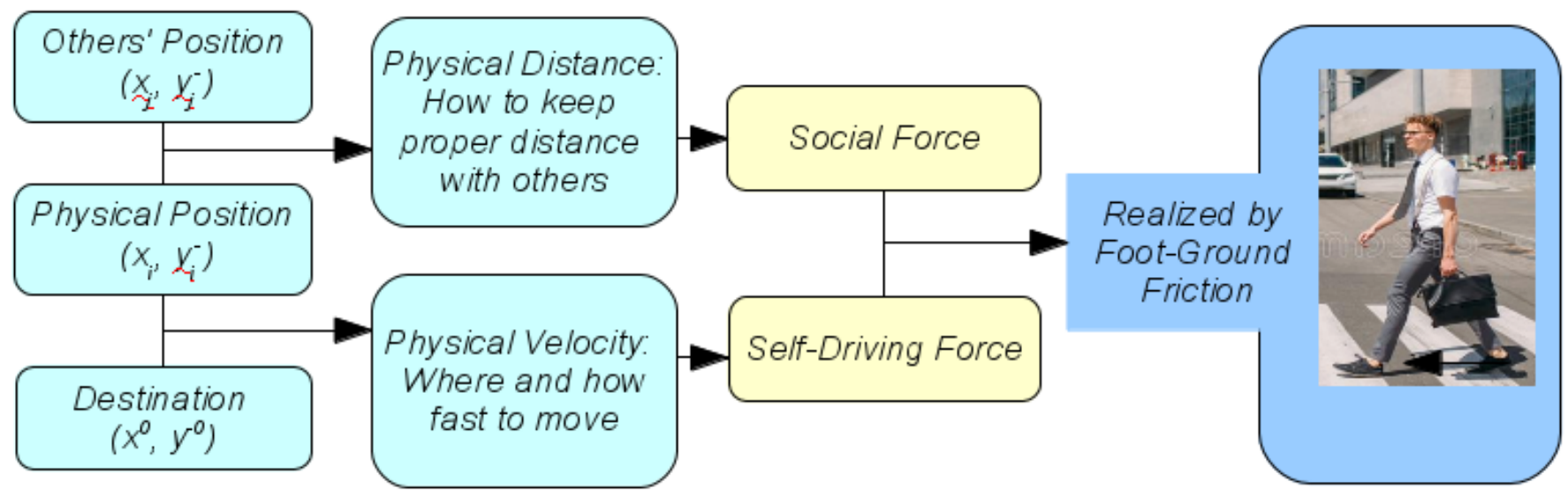

**Figure 2. Walking Behavior and Social-Force Model: When a human individual is walking, he will adjust the desired velocity and desired interpersonal distance such that he will move toward his destination while also keep a proper distance to surrounding people. Such perception and behavior is abstracted as the self-driving force and social force in the above mathematical model.**

## III. DYNAMICS OF PEDESTRIAN CROWD

In this section the microscopic model of individual movement will be aggregated to a macroscopic description of crowd movement. A fluid-like model is mainly presented to describe how psychological intention of people interacts with physical characteristics of motion (e.g., crowd speed and density). The fluid dynamics is first derived from force-based model in a general sense, and social force concept is next integrated. A major assumption of our analysis is high-density crowd, where crowd are assumed to be moving continuously on a planar surface, their movement can be considered as mass flowing with a specific rate in a two-dimensional space. The continuity hypothesis serves as a basis for the fluid-like model of pedestrians, and the specific characteristics of crowd motion include:

(a) Flow density and mass: The flow density is the number of pedestrians per area unit, and it is defined by $\rho=(\mathrm{d}N)/(\mathrm{d}x\mathrm{d}y)$, where $\mathrm{d}N$ is the number of pedestrians in the area of $\mathrm{d}x\mathrm{d}y$ (Figure 3). The crowd density characterizes the distance of people. Let $m_0$ denote the average individual mass in the crowd, and the mass of the flow in area of $\mathrm{d}x\mathrm{d}y$ is $m=m_0\rho\mathrm{d}x\mathrm{d}y$.

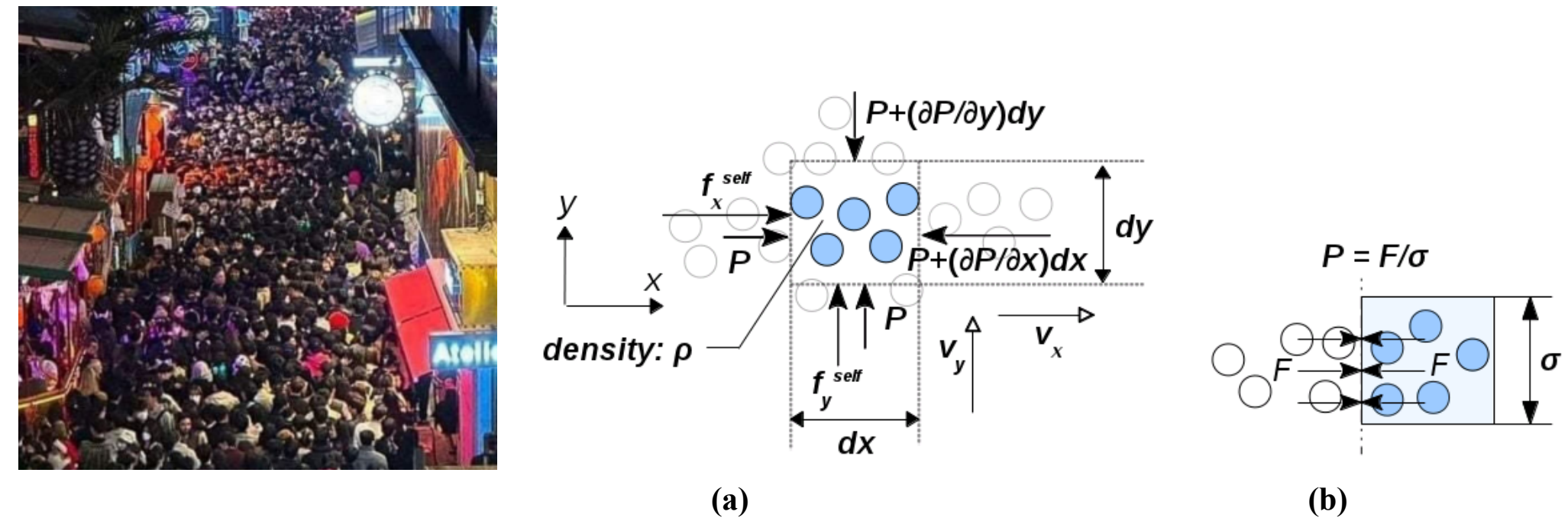

(c)

**Figure 3. Crowd movement along a passageway: As a crowd move on a planar surface, their movement can be considered as mass flowing with a specific rate in a two-dimensional space. The flow characteristics such as crowd speed and density are all functions of position ($x,y$) and time $t$, and Newton Laws is applied in such analysis.**

(b) Interactions: As people move collectively, their physical interactions of people are characterized by surface pressure $P$. It is a two-dimensional analog of common pressure concept in three-dimensional space, and it is the lateral force per unit length applied on a line perpendicular to the force, namely $P=F/\sigma$ as shown in Figure 3(c). (Discontinuity of $P$ in the crowd fluid)

(c) Physical motion: The velocity of the moving crowd is denoted by $\boldsymbol{v}$, and it can be decomposed orthogonally as $v_x$ and $y_y$, characterizing the moving speed along $x$ axis and $y$ axis, respectively (See Figure 3(b)).

(d) Motive Force: Human motion is self-driven. In physics, the motive force is commonly considered as frictions or pushing forces that people implement on the ground by feet. What differs creatures like human from non-creatures is that the motive force is intentionally generated by creatures such that they can decide where to go and how fast they move. Thus, the motive force is not only a physical concept, but also represents intentions of people in their mind. As a result, this paper presents the motive force by $\boldsymbol{f}^{\mathrm{self}}=m\boldsymbol{a}^{\mathrm{self}}$, where $\boldsymbol{a}^{\mathrm{self}}$ is called self-acceleration and it indicates intentions of people. In a similar way, $\boldsymbol{a}^{\mathrm{self}}$ can be decomposed as $a_{\mathrm{x}}^{\mathrm{self}}$ and $a_{\mathrm{y}}^{\mathrm{self}}$.

The flow characteristics as presented above (i.e., $\rho$, $P$, $v$, $a^{self}$) are all functions of position ($x,y$) and time $t$, and the crowd flow is unsteady flow in our analysis. With the assumption of the conservation of flow mass (i.e., mass continuity equation), we now study the moving crowd in the flow section of $\mathrm{d}x\mathrm{d}y$, where the mass is $m=m_0\rho\mathrm{d}x\mathrm{d}y$ . As people try to accelerate along the positive direction of $x$ axis, the resistance comes from the surrounding people, and $\mathrm{d}F$ thus becomes resistant to the crowd motion. The motive force is $\boldsymbol{f}^{\mathrm{self}}=m\boldsymbol{a}^{self}=\boldsymbol{a}^{self}m_0\rho\mathrm{d}x$. By the Newton Second Law we have

$$m_0\rho\left(\frac{\partial v}{\partial t}+(v\cdot\nabla)v\right)=a^{self}m_0\rho-\nabla P \tag{3}$$

Equation (3) corresponds to Euler's Equation in fluid mechanics and it demonstrates the Newton Second Law ($ma=\sum F$) for the crowd flow. Here we note that the derivation steps as above do not require any mathematical form of $\boldsymbol{a}^{self}$. In general, $\boldsymbol{a}^{self}$ should reflect the cognition process by which people perceive the physical world and form their goals in mind. Learning from the social-force model, this paper specifies the subjective targets in people's mind by desired velocity $\boldsymbol{v}^d$ and desired density $\rho^d$, and the self acceleration is given in a feedback manner as

$$
\begin{aligned}
a^{self} &= \left(a_v^{self} + a_\rho^{self}\right) \\
a_v^{self} &= k_1\left(v^d - v\right) = k_1\left[\left(v_x^d - v_x\right)i + \left(v_y^d - v_y\right)j\right] \\
a_\rho^{self} &= k_2\left(\rho^d - \rho\right)\nabla\rho = k_2\left(\rho^d - \rho\right)\left(\frac{\partial\rho}{\partial x}i + \frac{\partial\rho}{\partial y}j\right)
\end{aligned}
\tag{4}
$$

where $k1$ $k2$ are parameters that weigh differently on targets of $v^d$ and $\rho^d$, and they have specific units to form the acceleration. The desired speed $v^d$ and density $\rho^d$ represent the psychological target in people's mind, specifying the speed and interpersonal distance that people desire to realize. The physical speed $v$ and density $\rho$ indicate the physical characteristics that are being achieved in the reality. As a result, the difference $v^d$ -$v$ and $\rho^d$ -$\rho$ show the gap between the human subjective wish and realistic situation, and they form the motive force to make the physical variables approaching towards the psychological targets. In particular, $\boldsymbol{a}_v^{self}$ and $\boldsymbol{a}_\rho^{self}$ corresponds to the self-driving force and the social-force in the pedestrian model in Section 2.

Equation (3) and (4) should be jointly used with other equations (e.g., the conservation of mass; boundary conditions) in order to fully describe the flow characteristics in a given geometric setting. By plugging Equation (4) into Equation (3) and using orthogonal decomposition. Equation (3) and (4) can be rewritten as below with the equation of mass continuity.

$$
\begin{aligned}
m_0\rho\left(\frac{\partial v_x}{\partial t} + v_x\frac{\partial v_x}{\partial x} + v_y\frac{\partial v_x}{\partial y}\right) &= \left(k_1\left(v_x^d - v_x\right) + k_2\left(\rho^d - \rho\right)\frac{\partial\rho}{\partial x}\right)m_0\rho - \frac{\partial P}{\partial x} \\
m_0\rho\left(\frac{\partial v_y}{\partial t} + v_x\frac{\partial v_y}{\partial x} + v_y\frac{\partial v_y}{\partial y}\right) &= \left(k_1\left(v_y^d - v_y\right) + k_2\left(\rho^d - \rho\right)\frac{\partial\rho}{\partial y}\right)m_0\rho - \frac{\partial P}{\partial y} \\
\frac{\partial\rho}{\partial t} + \frac{\partial\left(\rho v_x\right)}{\partial x} + \frac{\partial\left(\rho v_y\right)}{\partial y} &= 0
\end{aligned}
\tag{5}
$$

From the perspective of psychology studies, the gap between the psychological desire (i.e., $v^d$ and $\rho^d$) and the physical reality (i.e., $v$ and $\rho$) relates to how much stress people will be experiencing (Staal, 2004). In other words, the difference $v^d$ -$v$ and $\rho^d$ -$\rho$ relates to the psychological concept of stress, and Equation (3) and (4) show that accumulation of such stress will motivate certain behavior of people. In particular, the gap of speeds characterizes the time-related stress, which relates to the time-pressure in psychological research. The gap of density reflects the a kind of space-related stress, indicating how people use space and interact with others in the collective movement. In particular, Equation (5) implies a feedback mechanism by which people's mind functions like a controller of their behavior: the targets in mind (i.e., $v^d$ and $\rho^d$) guide people to change their physical characteristics (i.e., $v$ and $\rho$), and such changes in the physical world are also feedback to people's mind. As a result, mind and reality interact in a closed-loop such that a balance can be reached between the psychological world of human mind and physical world of reality. Control theory can help to explore how $v$ and $\rho$ are changed given $v^d$ and $\rho^d$, and cognition science focuses on how people perceive the reality and thus adjust $v^d$ and $\rho^d$ based on their perceptions.

Table 1 On Conception of Stress in Self-Driven Crowd Fluid Model

| | *Opinion (Psychological Characteristics)* | *Behavior (Physic-Based Characteristics)* | *Difference between subjective opinion and objective reality* | *Forced-Based Term for Newton Second Law* |
|---|---|---|---|---|
| ***Time-Related Stress: Velocity*** | desired velocity $\boldsymbol{v}^d$ | actual velocity $\boldsymbol{v}$ | ***Time-Related Stress: Velocity*** $\boldsymbol{v}^d - \boldsymbol{v}$ or $\lvert\boldsymbol{v}^d\rvert/\lvert\boldsymbol{v}\rvert$ | ***Driving Force*** $\boldsymbol{f}^{drv} = k1\,(\boldsymbol{v}^d - \boldsymbol{v})$ |
| ***Space-related Stress: Distance*** | desired density $\rho^d$ | actual density $\rho$ | ***Space-related Stress: Distance*** $\rho^d - \rho$ or $\rho^d/\rho$ | ***Social Force*** $\boldsymbol{f}_{ij}^{soc} = k2(\rho^d - \rho)(i\partial\rho/\partial x + j\partial\rho/\partial y)$ |

How to determine $v^d$ and $\rho^d$ in practical computing is a challenging task. The direction of $v^d$ could be given by solving a Possion Equation, where each candidate destination of movement is a sink point, and the flow solver calculates the route as the crowd move to the sink (Korhonen, 2018; Korhonen and Hostikka, 2010). The computation result is a 2D velocity field that guides evacuees to the destination selected. The magnitudes of $v^d$ and $\rho^d$ are not easy to be given because they are inter-related issues. Suppose there are preset values given for $v^d$ and $\rho^d$, and in the simulation loop $v^d$ is first modified by any environmental stimuli that causes time-pressure, and $\rho^d$ is next adjusted by the difference $v^d$ -$v$. In other words, the time-related stress is the very initial cause of motion, and it is transformed into space-related stress for high-density crowd.

There is another problematic method to determine the magnitudes of $v^d$ in some existing research work. That is, the magnitude of $\boldsymbol{v}^{\mathrm{d}}$ is assumed to be a function of density such as Greenshields model $v^d = v_f(1-\rho/\rho_{jam})$, Greenberg model is $v^d = v_f \ln(\rho/\rho_{jam})$, Underwood model $v^d = v_f \exp(-\rho/\rho_{jam})$. These formula are partly obtained from the car-following problem, and all the models were extracted from empirical observation, suggesting that the physical speed reduces with increasing density $\rho$. The above formula are initially applied to the physical speed, not for desired speed. When physical speed and density evolve without desired velocity, the two variables could reach the equilibrium state such that the physical speed and density are balanced based on the empirical rules of Greenshields model and several others. However, a dynamic process with desired velocity $\boldsymbol{v}^{\mathrm{d}}$ is different, and it is a complex process involving the desired target $|\boldsymbol{v}_i^0(t)|$ in people's conscious mind. Thus, it is not reasonable to simply assume $|\boldsymbol{v}_i^0(t)|$ as a function of density. In sum our conclusion is that the physical speed in the equilibrium state is a function of density, but the desired speed is not.

## IV. ENERGY-BASED ANALYSIS OF PEDESTRIAN CROWD

By aggregating individual pedestrians to the flow of crowd, an energy balance equation is derived in this section to explore how energy is transformed from the psychological world of human mind to the physical world of universe. Such energy-based analysis helps to bridge a gap among psychological findings, pedestrian models and simulation results, and it further provides a new perspective to understand the paradoxical relationship of subjective wishes of human and objective result of their deeds in reality.

A scenario is discussed when people move through a passageway as shown in Figure 4. Here it is assumed that the width of the passageway is relatively small compared to the length of the crowd flow. The flow is thus assumed to be homogeneous in the direction perpendicular to the passageway direction (i.e., along $y$ axis in Figure 3(b)), and the flow characteristics vary only along the passageway direction (i.e., along $x$ axis in Figure 3(b)). In brief, this section focuses on the crowd movement in a passageway like in a one-dimensional space, and the possible physical forces from walls $f_W$ are assumed to be and equal in magnitude and opposite in direction (See Figure 4). As a result, it yields $\partial P/\partial y=0$, $\partial \rho/\partial y=0$, $\partial v_x/\partial y=0$, $v^d_y=0$ and $v_y=0$. In other words, $v=v_x$ , $v^d=v_x^d$ and $f^{self}=f_x^{self}$ and Equation (5) is thus rewritten as

$$m_0 \rho \left( \frac{\partial v}{\partial t} + v \frac{\partial v}{\partial x} \right) = \left( k_1 \left( v^d - v \right) + k_2 \left( \rho^d - \rho \right) \frac{\partial \rho}{\partial x} \right) m_0 \rho - \frac{\partial P}{\partial x}$$

$$\frac{\partial \rho}{\partial t} + \frac{\partial (\rho v)}{\partial x} = 0 \qquad (6)$$

**Figure 4. Crowd movement along a passageway: Equation (5) is simplified for one-dimensional analysis, describing how crowd speed, density and pressure are dynamically formulated in Euler's Equation.**

Mathematically, Equation (6) is to be jointly used with boundary conditions in order to fully describe the flow characteristics in a given geometric setting. Based on the existing theory in partial differential equations and control theory, the solution should be a wave function, which implies that the gap of $v^d$ -$v$ and $\rho^d$ -$\rho$ are periodic functions. In other words, $v$ and $\rho$ will sway around the target $v^d$ and $\rho^d$, and finally converge to the target value $v^d$ and $\rho^d$. If $v^d$ and $\rho^d$ are dynamically changed with time, then $v$ and $\rho$ will also track such changes with a time delay.

Further, when the crowd flow converges into the steady state, it implies $\partial v/\partial t=0$ and the Eulerian derivative becomes zero. For such steady flow, the flow characteristics (i.e., $\rho$, $F$, $v$, $a^{self}$) are time-invariant, and energy-based analysis is commonly applied. By taking the dot product with d$s$ – the element of moving distance – on both sides of Equation (6), an energy equation can be derived, which corresponds to the well-known Bernoulli equation in fluid mechanics. In particular, the element of distance in one-dimensional space is d$s$=d$x$, and thus we have

$$m_0 \rho v \frac{\partial v}{\partial x} \mathrm{d}x = a^{self} m_0 \rho \,\mathrm{d}x - \frac{\partial P}{\partial x} \mathrm{d}x \qquad m_0 v \,\mathrm{d}v = a^{self} m_0 \,\mathrm{d}x - \frac{\mathrm{d}P}{\rho} \tag{7}$$

Because the flow characteristics (i.e., $\rho$, $F$, $v$, $a^{self}$) are only functions of position $x$ for steady flow, it gives

$$m_0 \mathrm{d}\left(\frac{v^2}{2}\right) + \frac{\mathrm{d}P}{\rho} = a^{self} m_0 \,\mathrm{d}x \qquad \frac{m_0 v^2}{2} + \int_0^P \frac{\mathrm{d}P}{\rho} = \int_{x_0}^{x} m_0 a^{self} \,\mathrm{d}x + C \tag{8}$$

Here $m_0$ is the average individual mass and does not depend on moving speed $v$, and the equation can be integrated. The physical interactions are repulsive among people, and $P \geq 0$. Based on Equation (4) an energy balance equation is obtained as below

$$\frac{m_0 v^2}{2} + \int_0^P \frac{\mathrm{d}P}{\rho} = \int_{x_0}^{x} m_0 k_1 \left(v^d - v\right) \mathrm{d}x + \int_{x_0}^{x} m_0 k_2 \left(\rho^d - \rho\right) \frac{\partial \rho}{\partial x} \mathrm{d}x + C \tag{9}$$

Similar to the well-known Bernoulli Equation in fluid mechanics, Equation (9) can be interpreted by the principle of energy conservation, where the psychological drive can be considered as a special form of potential energy that arises from crowd opinion, and its behavioral manifestations are energy in physical forms.

The left side of Equation (9) includes energy in physical forms – kinetic energy and static energy. In addition, if the crowd movement is not horizontal, but on a slope, the gravity should be taken into account by including the gravitational potential $m_0gh$ on the left side of Equation (9), where $h$ is the altitude of the crowd position and $g$ is the gravitational acceleration. The kinetic energy is the common form that describes the energy regarding crowd motion. The static energy is an integral form that characterizes the physical interactions of people. The physical interaction comes into existence when the crowd density exceeds a certain limit. Therefore, $P$ can also be considered as a function of crowd density $\rho$. A mathematical expression of the static energy is exemplified as below. Let $\rho_0$ represent the crowd density when people start to have physical contact, and the interaction force is given by

$$P = K\eta\left(\rho - \rho_0\right) \qquad \int_0^P \frac{\mathrm{d}P}{\rho} = K\eta\left(ln\rho - \ln \rho_0\right) \tag{10}$$

where $K$ is a positive parameter. $\eta(\cdot)$ is a piecewise function such that $\eta(x)=0$ if $x<0$ and $\eta(x)=x$ if $x>=0$. The interaction force thus becomes nonzero when $\rho>\rho_0$ . The resulting static energy is given by Equation (10). In general, the moving crowd is a compressible flow: the crowd density varies in different places of the flow. Thus, the static energy is a function of density $\rho$. Moreover, the social contact in Helbing, Farkas, and Vicsek, 2000 is treated as an analog of a physical force. In contrast, social contact in our analysis is treated differently by introducing the desired crowd density $\rho^d$. The resulting static energy in Equation (10) only captures the physical interactions of people. The social contact is described in the self-energy. It can be repulsive or attractive, depending on the relationship of physical density $\rho$ and desired density $\rho^d$.

The right side of Equation (9) is energy involving crowd opinion, and it is called motivation energy or stress energy in this paper. This special kind of energy is expressed by an integral term, and it follows the typical formula that is the integral product of a force and moving distance along the direction of the force. The motivation energy in Equation (9) consists of two terms: one term drives people to adjust speed in a temporal space and the other one motivates people to adjust their social distance with others. This corresponds to the driving force and social force in the social force model. An interesting topic is that Equation (9) extends the law of energy conservation from the physical world of universe to the psychological world of human mind, implying potential transformation of motivation energy into certain physical form. In fact, the energy-based equation shows that the motivation energy functions like an energy depot (Ebeling and Schweitzer, 2001), and it arises in mind when people have desire doing something, and it will be ultimately transformed into certain physical energy in reality. In other words, energy in mind cannot vanish by itself, but must find an outlet to the physical world.

Table 2. Energy Balance in Crowd Movement

| Kinetic Energy | | Static Energy |
|---|---|---|
| $\frac{m_0 v^2}{2}$ | ⇦ ⇨ | $\int_0^P \frac{\mathrm{dP}}{\rho}$ |
| Influence ⇧ | | ⇩ Feedback |
| $\int_{x_0}^{x} m_0 k_1 (v^d - v) dx$ | ⇦ ⇨ | $\int_{x_0}^{x} m_0 k_2 (\rho^d - \rho) \frac{\partial \rho}{\partial x} dx$ |
| Self-Driving Characteristic | | Social Characteristic |

The motivation energy characterizes the collective opinion of people in mind. From the perspective of psychological principles, the variables of $v^d$ and $\rho^d$ thus have much freedom because they exist in people's mind. However, as people realize $v^d$ and $\rho^d$, their behavior are not free any more since certain realistic factors confine their deed in the physical world (e.g, the size of a passage may not permit all the people to move as fast as desired). If the physical variables reach the maximum while people still desire increasing them, the motivation energy will not be transformed to the physical forms as desired. In this situation, the stronger is the subjective wish in people's mind, the worse may become the situation in the reality, showing a paradoxical relationship of subjective wishes of human and objective result in reality. An example in this kind is the "faster-is-slower" effect as shown in Helbing, Farkas, and Vicsek, 2000.

## V. DOORWAY SCENARIO AND HEATING EFFECT

The energy-based analysis provides us a new perspective to reinterpret the faster-is-slower effect and Yerkes–Dodson law. If the motivation energy is transformed properly so that people are able to speed up, the psychological drive of motion will accelerate the crowd, and faster-is-faster effect shows up. In contrast, when the motivation energy cannot be transformed to the physical forms as desired, the faster-is-slower effect comes to being. Very importantly, the passage capacity determines the maximal amount of motivation energy that can be transformed to the physical forms, and it determines a critical threshold: below the threshold the psychological drive as expressed by the motivation energy is transformed to the kinetic energy and the crowd can accelerate as desired. Above the threshold the kinetic energy reaches the maximum, and excess of psychological drive will be transformed to the static form, resulting in an increase of crowd density. This answers the question about when the motivation energy is transformed to the kinetic form, and when to the static form.

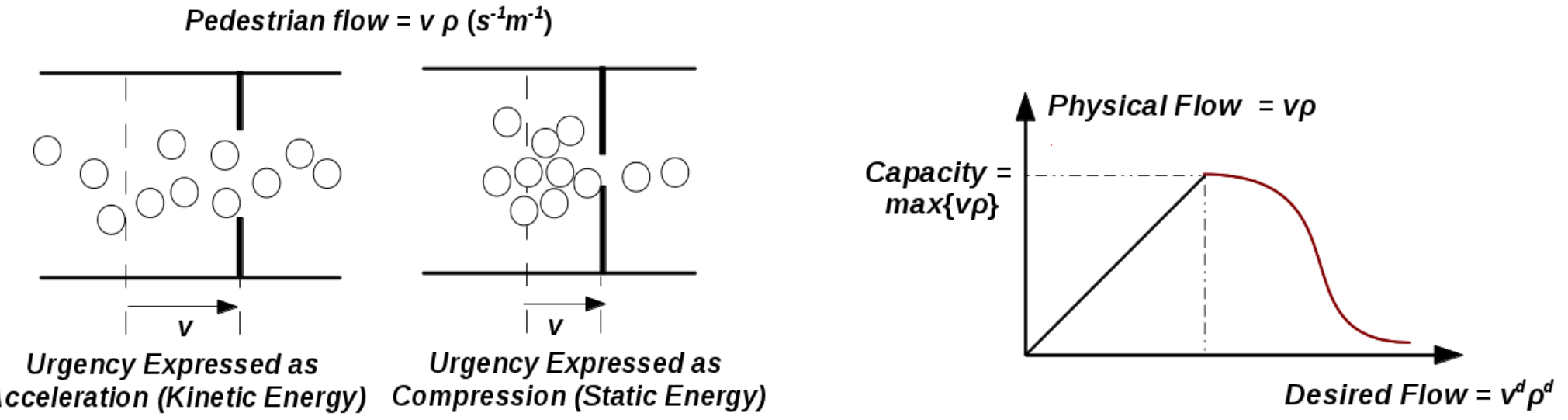

**Figure 5. Relationship of the desired flow $\rho^d v^d$ and physical flow $\rho$v: $\rho^d v^d$ indicates the collective demand of crowd movement and $\rho$v indicates the physical motion that the crowd realize. This figure reiterates the Yerkes-Dodson law at the macroscopic level, and $\rho^d v^d$ is the stress indicator, which represents the motivation level at the macroscopic level, and moderate stress improves the performance (i.e., speeding up crowd flow) while excessive stress impairs it (e.g., disordering and jamming).**

If the crowd density exceeds a certain limit, the pedestrian flow will decrease. Further increase of the crowd density could result in disorder events (e.g., jamming or stampeding). In this light our crowd flow model and energy-based analysis also reiterates the Yerkes–Dodson law: moderate stress improves the performance (i.e., speeding up crowd flow) while excessive stress impairs it (e.g., disordering and jamming). The relationship of the desired flow $\rho^d v^d$ and physical flow $\rho$v is plotted as shown in Figure 5, where $\rho^d v^d$ indicates the collective demand of crowd movement and $\rho$v indicates the physical motion that the crowd realize.

Table 3. Conservation of Energy with Fluctuation Effect and Gravitational Effect

<table>
<tr><td>Kinetic Energy</td><td>$\frac{m_0 v^2}{2}$</td><td rowspan="4">Energy<br>in Physical World<br>↑↓<br>Energy<br>from Conscious Mind</td></tr>
<tr><td>Static Energy</td><td>$\int_0^P \frac{dP}{\rho}$</td></tr>
<tr><td>Gravitational Energy</td><td>$m_0 g\hbar$</td></tr>
<tr><td>Motivation & Fluctuation<br>(Psychological Features in Collective Opinions)</td><td>$\int_{x_0}^{x} m_0 a^{self}\,dx+\int_{x_0}^{x} m_0 a^{\xi}\,dx$</td></tr>
</table>

Continuing with the above energy-based analysis we will next take the gravitational force and fluctuation random force into account, and explain the risk of "down-slope" and "heating effect" in crowd egress. Down-slopes are commonly known as stairways, where the gravity is effective. As a result, the gravitational potential is necessarily taken into account in Equation (9), and it can either facilitate or impede acceleration or compression of a crowd. When people desire moving faster ( $v^d>v$), the gravity facilitates acceleration on a down-slope while resists such acceleration on an up-slope. A similar effect is also for compression. Because people usually move downstairs in egress, acceleration and compression will be more intensive in a down-slope, and the risk of disorder and blocking is thus increased.

The heating effect was known for the "heating-by-freezing" in Helbing et al., 2002. Such a heating effect implies that the fluctuation force in Equation (1) is in higher magnitude (Helbing et al., 2002). From the perspective of energy-based analysis people are often more emotional when watching sports games, attending concert or joining parties, and they usually exhibit more energy in these occasions. Even if they are not in physical motion, they will release such energy by shouting or other behavior. So stadium, concert or nightclubs are a kind of public places where people's inner energy will find an outlet to the physical world, and such effect on crowd is similar to heating process as mentioned in Helbing et al., 2002. Corresponding to the fluctuation force in Equation (1), a fluctuation acceleration is added in Equation (11), and it indicates the heating effect on the crowd. For heating effect the fluctuation force is not toward a certain direction and it represents irrational mind of people. In contrast the desired velocities and desired distance are deterministic in directions and represent rational mind of people.

$$\frac{m_0 v^2}{2}+\int_0^P \frac{\mathrm{dP}}{\rho}+m_0 g\hbar=\int_{x_0}^{x} m_0 a^{self}\,dx+\int_{x_0}^{x} m_0 a^{\xi}\,dx+C \qquad (11)$$

When the "heated" crowd are further compressed at a downstairs bottleneck in egress, the risk of disorder and stampede is much higher than in other places. In a list of historical events of stampeding in Still, 2016, it suggests that certain gathering places such as stadium are more frequent to occur stampede in emergency egress. Similar evidence is also listed in Helbing et. al., 2002, and one of the crucial reasons is that bottlenecks in a stadium are often not horizontal, but on stairways (See Figure 6). Another important reason is that people usually feel excited when watching football games, and the crowd are thus "heated" as Helbing described. From the practical viewpoint, it is almost impossible to clam down the crowd who join a sport event in stadium, but it is feasible to design egress facilities such that the bottleneck (e.g, a narrow passageway) are not placed on down-slope areas. In other words, it is better to design the downstairs ways in a wide or open area, and when people get gathered at the bottleneck, the passageway should be horizontal.

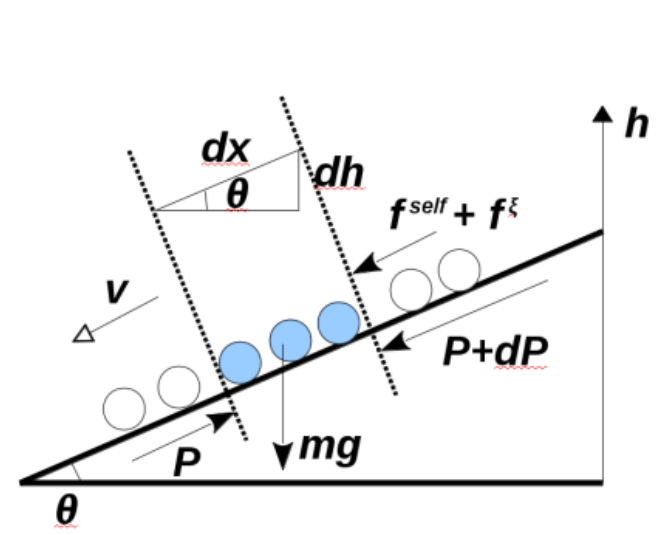

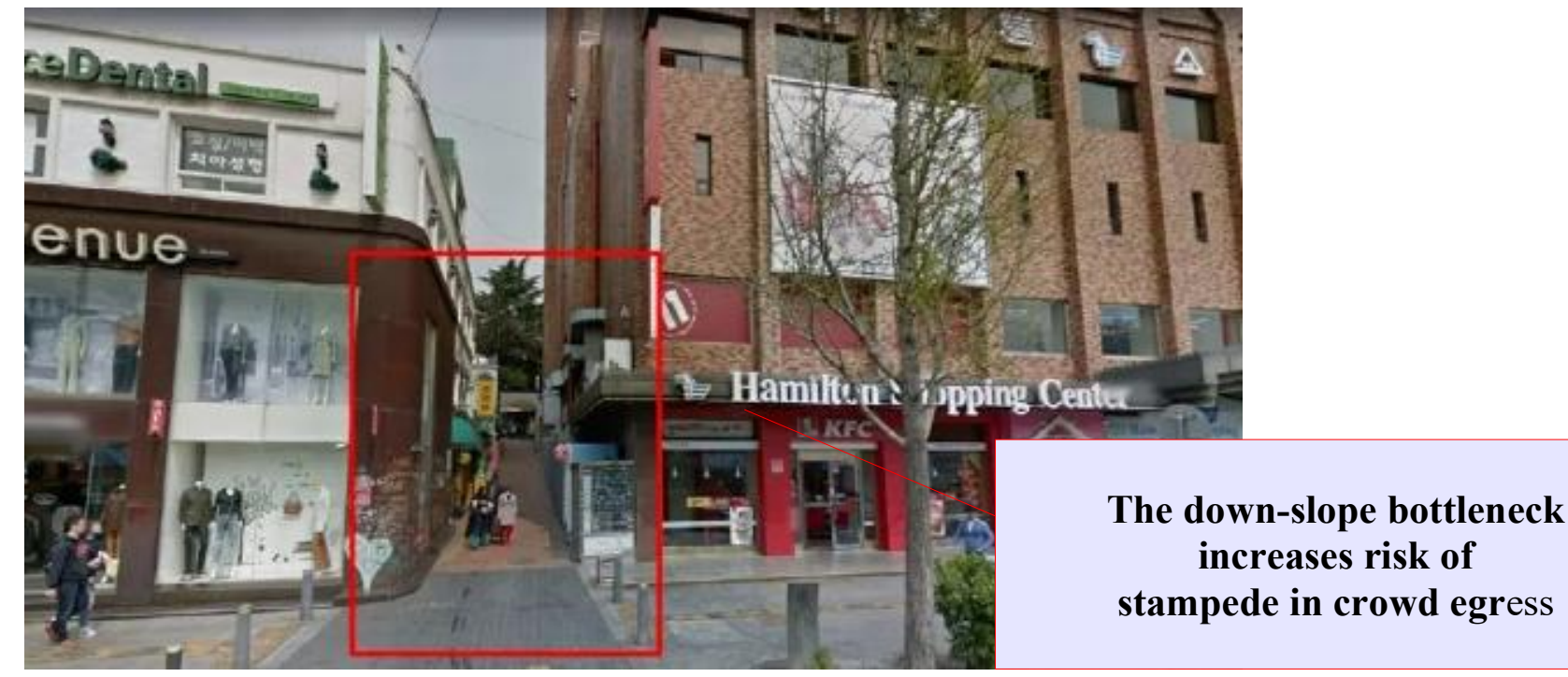

**Figure 6. A Bottleneck Passageway at Itaewon: When the "heated" crowd are further compressed at a downstairs bottleneck in egress, the risk of disorder and stampede is much higher than in other places. In a list of historical events of stampeding in Still, 2016, it suggests that certain gathering places in festivals are more frequent to occur stampede than other normal places.**

## I. CONCLUSIONS

Human stampedes occurred at various types of gatherings around the world. Since 1990s, there were an estimated over 50 crowd surge incidents, which led to over 6900 deaths and many injuries. Surprisingly, a disorderly crowd often turn into a fatal hazard, not because of any external threats such as fire or explosion, but by the crowd itself. In a sense, human crowd are self-organized into an deadly disaster unexpectedly and unconsciously, and certain emotional factors play a vital role such as stress and impatience.

To investigate the major cause of such crowd disasters, an energy balance equation is derived in this paper to explore how different forms of energy are transformed and balanced during crowd surge incidents. Such energy-based analysis advances cross-disciplinary knowledge of psychology, crowd modeling and engineering studies, and it also provides a new perspective to mathematically describe the paradoxical relationship of human subjective opinion and objective result of their deeds in reality, such as the "faster-is-slower" effect during crowd disasters. As an integration of multiple disciplines, our fluid-based analysis holds promise of long-term impact on crowd safety management by exploring transformative optimization methods for crowd gathering events.

## ACKNOWLEDGMENTS

This paper is especially dedicated to Prof. Peter Luh for memorizing his insightful guidance in this study subject. The authors are sincerely thankful to Prof. Kerry Marsh, Xiaoda Wang, Neal Olderman, Shi-Chung Chang, Bo Xiong for helpful discussion. The authors appreciate the research program funded by NSF Grant # CMMI-1000495 (NSF Program Name: Building Emergency Evacuation - Innovative Modeling and Optimization)..